\documentstyle[prb,aps,psfig]{revtex}
\begin{document}
\draft
\title{The specific heat and optical birefringence of $Fe_{0.25}Zn_{0.75}F_2$}
\author{W. C. Barber and D. P. Belanger}
\address{Department of Physics, University of California\\
Santa Cruz, California 95064}
\date{\today}
\maketitle
\begin{abstract}

The specific heat $(C_m)$ and optical birefringence $(\Delta n)$ for the 
magnetic percolation threshold system $Fe_{0.25}Zn_{0.75}F_2$ are
analyzed with the aid of Monte Carlo (MC) simulations.  Both $\Delta n$
and the magnetic energy $(U_m)$ are
governed by a linear combination of near-neighbor spin-spin correlations,
which we have determined for $\Delta n$ using MC simulations modeled closely
after the real system.  Near a phase transition or when only one
interaction dominates, the temperature derivative of the birefringence
$[{d(\Delta n)}/{dT}]$ is expected to be proportional
$C_m$ since all relevant correlations necessarily have the same temperature
dependence.  Such a proportionality does not hold for $Fe_{0.25}Zn_{0.75}F_2$
at low temperatures, however, indicating that neither condition
above holds.  MC results for this percolation system
demonstrate that the shape of the temperature derivative of correlations
associated with the frustrating third-nearest-neighbor interaction differs
from that of the dominant second-nearest-neighbor interaction, accurately
explaining the experimentally observed behavior quantitatively.
\end{abstract}

\pacs{75.40.-s,75.50.Ee,75.50.Lk}

Measuring the linear optical birefringence $(\Delta n)$ in
anisotropic, antiferromagnetic
crystals undergoing magnetic phase transitions is a powerful way of determining
the magnetic specific heat ($C_m$) critical behavior.  It has been
shown\cite{j73,fg84,bkj84} that the temperature derivative of the
optical birefringence $[d(\Delta n)/dT]$ is proportional to $C_m$.  In
many cases, the birefringence technique has provided the most precise
experimental determinations
of universal critical behavior parameters\cite{by91} in pure and randomly
mixed and dilute magnetic Ising systems.  For the case of the
three-dimensional $(d=3)$ random-field Ising model (RFIM), which applies
for a dilute antiferromagnet with an applied field along the ordering
direction, birefringence measurements yielded evidence of a
phase transition\cite{bkjc83}.  The advantages of the technique are
threefold:  the technique is typically much easier to employ than traditional
heat pulse techniques; the effects of concentration gradients inevitably
present in mixed and dilute crystals can be greatly reduced\cite{bkj84}; and
the phonon
contributions to the specific heat are greatly suppressed in the birefringence.
Since the transition typically varies with concentration in mixed and dilute
systems, the critical behavior is often masked by the concentration gradients
quenched into the system during growth.  The laser beam used in the optical
technique can be aligned perpendicular to the gradient, often reducing the
gradient effects by an order of magnitude.  This has been crucial in the
study of random-exchange Ising model (REIM) and RFIM systems in $d=2$ and $3$.
The virtual elimination
of the phonon background has allowed detailed analysis of the specific heat
in $d=1$ and $2$ systems.  For $d=2$ this has allowed a detailed
comparison\cite{nbkji83} of the magnetic specific heat of the pure system
and the Onsager solution to the $d=2$ Ising model and, for dilute systems,
a scaling analysis of the destruction of the phase transition by random fields
in the $d=2$ Ising model\cite{fkjcg83}.  For $d=1$, the technique has been used
successfully to determine the exchange constants in systems where the phonon
background overwhelms the magnetic contributions to conventional specific
heat data\cite{jlfagy96}.

That $C_m$ and $d(\Delta n)/dT$ are proportional in
certain cases is a consequence of the
fact that both $\Delta n$ and the magnetic energy $(U_m)$
are governed by a linear combination of near-neighbor spin-spin
correlations~\cite{fg84}.  This proportionality has been experimentally shown 
explicitly\cite{bnkjlb83} for the $d=3$ pure Ising system $FeF_2$
as well as its site-diluted counterparts\cite{sb98,db89}
$Fe_{0.93}Zn_{0.07}F_2$ and $Fe_{0.46}Zn_{0.54}F_2$ by comparing
directly the pulsed heat and optical data.  This proportionality has been
demonstrated both in zero applied field ($H=0$) and in applied fields.
The magnetically dilute crystals of $Fe_{1-x}Zn_{x}F_2$
have been the most extensively studied\cite{by91} realizations of
the RFIM, an important class of systems
with randomness and frustration.  Such experimental verification has
been crucial in this controversial field since some authors\cite{comment}
have expressed reservations about the proportionality for $H>0$,
i.e.\ the RFIM case.

Despite the experimental verification of the proportionality between
$C_m$ and $d(\Delta n)/dT$, it is important to verify the details of
the mechanism for the proportionality near phase transitions in a
specific case.
One opportunity to analyze the detailed relationship between $C_m$ and
$d(\Delta n)/dT$ presents itself in the compound $Fe_{0.25}Zn_{0.75}F_2$,
which, with respect to the dominant exchange interaction, $J_2$, is at
the magnetic percolation threshold, the concentration below
which no ordering is possible.  Throughout the rest of this work we
will refer to the percolation threshold concentration as the
appropriate one if only $J_2$ were being considered.  The presence of
other interactions
make the concept of percolation more complicated and they can play
an important role in the ordering processes even when they are very small.
Interestingly, $d(\Delta n)/dT$ data for $Fe_{0.25}Zn_{0.75}F_2$
exhibit a change\cite{thesis} in sign near
$T=8$~K.  While the high temperature behavior of $d(\Delta n)/dT$ is
accurately proportional to $C_m$, the sign reversal in $d(\Delta n)/dT$
at low temperature is not reflected, of course, by the behavior of $C_m$.
We present data obtained from both the optical and pulsed heat techniques
that show the nonproportionality between $C_m$ and $d(\Delta n)/dT$
at low $T$, as shown in Fig.\ 1.
We also present Monte Carlo (MC) simulation results that can be used to
explain why the proportionality breaks down between $C_m$ and
$d(\Delta n)/dT$ at low temperatures.  $U_m$ and $\Delta n$ each depend on the
near-neighbor spin-spin correlations, but with different weights.  We show that
whereas $U_m$ is dominated by the second-nearest-neighbor correlation,
$\Delta n$ is predominantly governed by the
third-nearest-neighbor correlation that shows a different temperature dependence
below $15$~K.  The results of this study are consistent with theory regarding
the proportionality between $C_m$ and $d(\Delta n)/dT$ near a phase transition,
where the different spin-spin correlations must necessarily have the
same temperature dependence.   We also demonstrate that
an applied field has a predictable effect on both $U_m$ and $\Delta n$.

The compound $Fe_{1-x}Zn_{x}F_2$ is ideal for RFIM studies. 
Pure $FeF_2$ is well modeled by the Hamiltonian
\begin{equation}
{\cal H} = \sum_l\sum_{i<j}J_l\vec{S_i}\vec{S_j} + D\sum_{i}\vec{S_i}^2 \quad .
\end{equation}
The $\vec{S}=2$ system has the interaction strengths $J_{1}=-0.069$~K,
$J_{2}=5.278$~K, and $J_{3}=0.279$~K, as determined from neutron inelastic
scattering measurements\cite{hrg70}.  The large crystal-field anisotropy
$D=9.29$~K
persists while the exchange interactions remain largely constant as the
magnetic spins are diluted\cite{a80} in  $Fe_{1-x}Zn_{x}F_2$, making this
an excellent Ising system for all magnetic concentrations $x$. Single
crystals can be grown for all $x$ with very 
small concentration gradients~\cite{kfjb88} and with superb structural quality.
The magnetic interactions are dominated by the antiferromagnetic 
second-nearest-neighbor super-exchange interaction, $J_{2}$,
between the body-center and body-corner ions.
All other interactions are negligible except near the percolation threshold
concentration, where a small frustration due to the third-nearest-neighbor
interaction between ions of the same sub-lattice along the direction
perpendicular to the spin-ordering $c$-axis,
$J_{3}$, becomes important\cite{syp79}.
At low temperatures, the Heisenberg character of the spin-spin interaction
is less important and we might expect $U_m$ to be fairly represented,
for $H=0$, by the REIM Hamiltonian,
\begin{equation}
{\cal H} = \sum_l\sum_{i<j>}J_l\epsilon _i \epsilon _j S_iS_j \quad ,
\end{equation}
where $S_{i}=\pm 2$, $\epsilon _i =1$ if site $i$ is occupied and zero
otherwise, and $J_l$ is the strength of the $l^{\rm th}$ nearest-neighbor
interaction.  $C_m$ is then given by a sum of temperature derivatives
of spin-spin correlations,
\begin{equation}
C_m = \frac{d(U_{m})}{dT} = \sum_{l}\sum_{i<j}J_{l}\epsilon _i \epsilon _j\frac{d(<S_{i}S_{j}>_l)}{dT} \quad ,
\end{equation}
where the $J_l$ are assumed independent of temperature.
The temperature dependent birefringence $(\Delta n)$, the difference between the indices of
refraction along the spin ordering 
axis and perpendicular to it, depends only on a linear combination of the
same correlation functions.  The temperature derivative yields
\begin{equation}
\frac{d(\Delta n)}{dT} = \sum_{l}\sum_{i<j}I_{l}\epsilon _i \epsilon _j\frac{d(<S_{i}S_{j}>_l)}{dT} \quad .
\end{equation}
In general the values of the coefficients $I_l$ associated with $\Delta n$
are not related to the respective values $J_l$.
Nevertheless, we will find a proportionality between $C_m$ and
${d(\Delta n)}/{dT}$ in a temperature regime where
the temperature dependence of the contributing correlations $<S_{i}S_{j}>_l$
are all proportional.  This is necessarily the case in the critical region of a phase
transition where the correlation length grows larger than any relevant
interaction length and all spin-spin correlations necessarily have the
same temperature dependence.  We may also find an excellent proportionality
between $C_m$ and $\Delta n$ in the case where one magnetic interaction
dominates in both $C_m$ and ${d(\Delta n)}/{dT}$.
Such is the case\cite{bkj84} of $MnF_2$, where the proportionality holds
accurately over a very wide temperature range, $5<T<100$~K.
A lack of proportionality has been observed in
pseudo-low-dimensional systems\cite{fg84,ll88} where neither
of the above conditions holds.  The percolation threshold
presents another
opportunity to observe a breakdown in proportionality between $C_m$ and
${d(\Delta n)}/{dT}$, since small frustrating interactions can affect
spins with few or no neighbors connected by the dominant exchange interaction.
We will show that this is indeed the case for the percolation threshold sample 
$Fe_{0.25}Zn_{0.75}F_2$.  We will also show that the behavior is readily
interpreted using the results of MC simulations of this
Ising antiferromagnet.

The $Fe_{0.25}Zn_{0.75}F_2$ crystal was cut from a boule 
grown~\cite{nk} at the University of California, Santa Barbara.
For the adiabatic heat pulse technique, the $2.24$~g sample was mounted
on a thin sapphire plate using GE7031 varnish.  A small Stablohm 800 wire
heater, connected to a four-wire constant-power supply, was wound and varnished
onto the sapphire plate.  A shielded carbon thermometer
was attached with varnish to the sample and was connected
using a four-wire technique to a current ratio transformer resistance
bridge.  The sample was suspended inside a sample chamber by $0.0254$~mm
Be-Cu wires used to connect to the thermometer and heater.  While providing good
electrical connections, the alloy wires provide only a small heat
leak from the sample to the copper sample chamber.  In this way,
the heat generated by the thermometer leaks away from the sample
in a way controlled by a temperature difference, $\delta T$,
between the sample chamber and the sample.  In the
absence of a heat pulse, the sample temperature can remain constant
if $\delta T$ is properly controlled.  This is done by controlling
$\delta T$ with a bridge and controller using the sample thermometer and
a thermometer located in a cold finger, commercially calibrated for $H=0$,
upon which the sample chamber is mounted.  A second thermometer in the
cold finger is used with a second bridge to determine the absolute temperature.
A preliminary calibration is performed to determine $\delta T$ versus $T$
such that the sample temperature does not drift in the absence of a heat
pulse.  The calibration is incorporated
into computer control algorithms so that the balance is automatically
preserved over the entire temperature range during a specific heat
experiment.  The computer applies a pulse and determines the resulting
change in temperature, from which the specific heat is calculated.
The thermometry sensitivity is approximately $50$~$\mu$K.
The pulse energy ranged from approximately $3$~$\mu$J at the lowest temperatures
to $8$~mJ at the highest temperatures where measurements were made.
After collecting specific heat data for the sample and addenda, the sample
is removed from the sample chamber and the thermometer is fixed to the sapphire
plate.  The specific heat of the thermometer, the sapphire plate, the varnish
and the wires is then measured at $H=0$. This background specific heat is
subtracted from all of the specific heat data before further analysis.
The phonon contribution to the total specific heat is approximated by the 
Debye model, valid at temperatures small relative to the Debye temperature.
The Debye temperature\cite{kh64} for pure non-magnetic $ZnF_2$ is $250$~K
and that of $FeF_2$ is $256$~K.  We weighted the two pure Debye temperatures
by the respective concentrations to obtain a Debye temperature of $252$~K
for $Fe_{0.25}Zn_{0.75}F_2$.  The Debye specific heat is calculated with
this Debye temperature and is subtracted from the data 
leaving the magnetic component of the specific heat ($C_m$) which can
then be used in comparisons with $\Delta n$ and MC results.

For the birefringence $(\Delta n)$ measurements,
two faces were polished parallel to the spin-ordering $c$-axis.
A linearly polarized laser beam ($\lambda = 632.8$~nm) impinges normally upon
the sample with a polarization $45^{\circ}$ to the $c$-axis.  The beam
traverses a distance of $8.05$~mm through the sample.  A $0.5$~mm pin-hole
in front of the sample minimizes the effects of concentration gradients
and vibrations.  The S\'{e}narmont technique is used
to measure $\Delta n$ with a resolution of
$2 \times 10^{-9}$.  Calibrated carbon resistance thermometers,
used for their low field dependence and high sensitivity, yield
a temperature stability better than $50$~$\mu$K. 
The sample was zero-field-cooled (ZFC) to $5$~K before slowly raising 
the temperature in approximately $0.1$~K steps.  About $400$~s were required
to establish equilibrium and to measure $\Delta n$ at each step.
Reasonable variations in the rates of heating and cooling and stabilization
times had no observable effect on the data.

The correlation functions for the first three nearest-neighbor pairs have been
calculated using MC simulations.
The magnetic lattice
corresponding to the body-centered-tetragonal $Fe_{1-x}Zn_{x}F_2$ lattice
is described as two cubic
sub-lattices of size $L \times L \times L$ each, delineated as one dimensional
arrays bit coded to accommodate large lattice sizes.
All of the results reported here were obtained with
$L=256$, corresponding to more than $3.3 \times 10^{7}$ sites
magnetically occupied with probability $x$. 
Beginning at high temperature each magnetic site
is randomly visited many times and flipped with a probability given by either
the heat bath or metropolis algorithm in temperature steps of $0.01$~K. Both
periodic boundary conditions and free boundary conditions were
applied to the lattices as they were cooled and then warmed in magnetic fields
of $H  = 0$ and $2$~T. The first three nearest-neighbor interactions were included
in the Hamiltonian with values taken from spin-wave dispersion 
measurements\cite{hrg70} for $FeF_{2}$, where $J_{1}=-0.069$~K, $J_{2}=5.278$~K,
and $J_{3}=0.279$~K. The correlation functions for these three types of
neighbors were calculated at each temperature step and averaged over ten
different configurations of magnetic spins. Increasing the number of
MC steps per temperature step dramatically did not change the results once a
minimum number of steps is performed so that the simulation stays in
quasi-static equilibrium at each temperature.  The proportionality constant
between $C_m$ and ${d(\Delta n)}/{dT}$ is found\cite{sb98,db89}
to be the same in $Fe_{0.93}Zn_{0.07}F_2$ and $Fe_{0.46}Zn_{0.54}F_2$ such that,
$A C_m$ = ${d(\Delta n)}/{dT}$, where $A=9.17 \times 10^{-6}$.  Our
sample $Fe_{0.25}Zn_{0.75}F_2$ is near the percolation threshold concentration
and cannot obtain long-range order associated with $J_2$.
However there is short-range order which produces a rounded peak in the $C_m$
data at about $T=23$~K.  ${d(\Delta n)}/{dT}$ vs.\ $T$ data agree with the $C_m$
data near this peak when the proportionality constant $A=9.17 \times 10^{-6}$
is used as shown in Fig.\ 1.  However, at low temperature $C_m$ and
${d(\Delta n)}/{dT}$ are no longer proportional.

MC simulations provide the opportunity to investigate the
relative importance of the different nearest-neighbor interactions on
the birefringence.  The second-nearest-neighbor interaction dominates the
energy Hamiltonian.  Therefore the specific heat is dominated
by the temperature derivative of the second-nearest-neighbor correlation
at the percolation threshold concentration.  However the simulations indicate
that $d(\Delta n)/dT$ is significantly governed by the temperature derivative
of the third nearest-neighbor correlation.  This is evident because the
two different correlations have contrasting low temperature behaviors, with
only the third-nearest-neighbor one changing the sign of its slope at
low temperature as shown in Fig.\ 2.  

In Fig.\ 3 we show the MC results for the Ising $C_m$ in the
percolation threshold concentration simulation along with the
$d=1$ Ising and Heisenberg exact models.  It is reasonable to expect that
the $d=1$ Ising model should correspond roughly to the behavior of
the simulation.  In both systems the spins have an average of two
neighbors, there is a similar energy gap and each should have a vanishing
specific heat at high temperature.  The $d=1$ Heisenberg model, on the
other hand, has a finite specific heat at zero temperature (the model
is not strictly followed by any real system at low temperature). 
This comparison does help to explain the fact that the MC data, shown
again in Fig.\ 4 as the small amplitude curve, do
not correspond very well to the experimental data.  Although
the real system has a large anisotropy, it nevertheless has a Heisenberg
interaction.  While the system exhibits an energy gap and its specific
heat falls to zero as the temperature is decreased, a significant
amount of specific heat near $T=10$~K is attributable to this Heisenberg
interaction.  Since we are interested in the relative contributions to
the biregringence from the frustrating third nearest interaction,
we choose to multiply the pure Ising behavior of the MC data by
a factor of $1.5$, thereby obtaining the good fit for $T<20$~K
exhibited in Fig.\ 4.

As mentioned above, at magnetic concentrations $x=0.93$ and
$x=0.46$ it has been shown\cite{sb98,db89} that the proportionality between
$C_m$ and $d(\Delta n)/dT$ is approximately the same as for pure $FeF_2$.
If we assume the same proportionality for the percolation threshold concentration
in the temperature regime where $C_m$ and $d(\Delta n)/dT$
are proportional and that the third nearest-neighbor correlation
dominates the behavior of the birefringence, we obtain
\begin{equation}
A {J_2}\sum_{i<j}\epsilon_i\epsilon_j\frac{d(<S_{i}S_{j}>_2)}{dT} = {I_3}\sum_{i<j}\epsilon_i\epsilon_j\frac{d(<S_{i}S_{j}>_3)}{dT} \quad .
\end{equation}
Furthermore, in this temperature regime we expect
\begin{equation}
\sum_{i<j}\epsilon_i\epsilon_j<S_{i}S_{j}>_2 = 2 \sum_{i<j}\epsilon_i\epsilon_j<S_{i}S_{j}>_3 \quad ,
\end{equation}
where the factor of two arises because there are eight second-nearest
neighbors and only four third-nearest neighbors.
This allows us to calculate $I_3$ in the previous equation to be
$4.85 \times 10^{-5}$.  With this value of
$I_3$ and the third nearest-neighbor correlation from MC
simulations we obtain a good fit to ${d(\Delta n)}/{dT}$ at low
temperature using the same factor of 1.5 used for $C_m$, as shown
in Fig.\ 5.  We stress that we have calculated $I_3$ in a regime where
$C_m$ and ${d(\Delta n)}/{dT}$
are proportional and used this value to fit the simulation results to
${d(\Delta n)}/{dT}$, so it is striking that the fit works well below $T=8$~K
where $C_m$ and ${d(\Delta n)}/{dT}$ are no longer proportional.
This provides strong evidence that the third
nearest-neighbor correlation dominates the birefringence
at low temperature and in the region where $C_m$ and $d(\Delta n)/dT$
are proportional.  However a small contribution from other correlations can
not be ruled out.  Detailed fits
between the correlations calculated by MC simulation and the
birefringence below $8$~K suggest such contributions
are less than ten percent.

To understand why the small, frustrating third-nearest-neighbor interaction
affects the third-nearest-neighbor correlation but has little effect on
the second-nearest-neighbor correlation, we turn to the
computer simulations.  When the simulation is run with only the $J_2$ interaction 
included in the Hamiltonian, all three nearest-neighbor correlations
have the same temperature dependence as expected. 
Close inspection of the MC simulation result show that 1.60\% of the sites
flip at low temperature as a result of the frustration.
This is close to the probability of having a site that contains a spin
having $J_3$ interactions but no $J_2$ interactions which is 1.71\%.  This
suggests that it is the spins that are not dominated by the $J_2$ interactions
that can order via the $J_3$ interaction.  The ordering of these spins
is thereby dominated by the $J_3$ interaction and the temperature
derivative of the difference between third nearest-neighbor correlations
calculated with and without the $J_3$ interaction present
shows a typical Schottky peak around a
temperature comparable to the interaction strength times average
number of neighbors. 
Since the probability of finding such spins decreases dramatically
above the percolation threshold concentration, no such effect should be
observed at larger concentrations.

We finally turn to the effect of an applied field on the behavior
of $d(\Delta n)/dT$ and, hence, the third-nearest-neighbor correlations.  
Having determined the zero-field proportionality between the
birefringence data and simulations of the $J_3$ correlations at low
temperature, we can now test whether the same proportionality
holds in an applied field even though the correlations themselves
change significantly with an applied field at low temperatures.
In Fig.\ 5 we show the measured behavior of $d(\Delta n)/dT$ vs.\ $T$
in an applied field $H=2$~T along the $c$-axis.  Also shown in the figure
is the MC prediction for the behavior based on the value of $I_3$
obtained for zero field.  The fit is clearly good.  Hence, the value of
$I_3$ has not changed significantly and the simulation correctly
describes the temperature dependence of the birefrincence in the
presence of an applied field.  This result is important since
the birefringence technique has been used extensively\cite{by91} to
study the RFIM which is realized by applying
a field to dilute anisotropic antiferromagnets above the percolation
threshold concentration.  Near a phase transition the proportionality
between $C_m$ and $d(\Delta n)/dT$ holds in a field and the birefringence
technique will yield the proper critical behavior.  This is important
because the birefringence technique invariably yields higher quality
data.

We have demonstrated that a small frustrating interaction, 
$J_3$, affects the corresponding correlation,
causing a reversal in the sign of the temperature derivative at temperatures
comparable to the interaction strength $J_3$.  In turn, this causes a
reversal in the temperature derivative of the birefringence in the
case of $Fe_{0.25}Zn_{0.75}F_2$.
The behavior of the birefringence in this system is fortuitious
because it allows us to determine that third-nearest-neighbor correlation
is the dominant one responsible for the birefringence.  In most other
cases of $d=3$ systems, one cannot determine the source of the
birefringence because all the correlations have similar temperature
dependences.  The fact that the proportionality, given by $A$, is nearly
constant for $x$ between the percolation threshold and pure
concentrations in $Fe_xZn_{1-x}F_2$ strongly suggests that
the birefringence is dominated by the third-nearest-neighbor
correlations for all concentrations.
We used the value of $J_3$ measured for pure $FeF_2$.  The good agreement
between the MC simulation and the data suggests that the strength of
the frustrating interaction $J_3$ is close to the same value
in the percolation threshold concentration sample.
Spin-glass-like behavior\cite{mrc91} near the percolation threshold has
strongly hinted at frustration in $Fe_{0.25}Zn_{0.75}F_2$, but the
strength of the frustration has not been previously determined.
Far above the percolation threshold concentration, or at high temperatures near 
percolation, this frustrating interaction has little affect on the
ordering processes or correlations, which all have the same
temperature dependence.

By comparing experimental data for $C_m$ and $d(\Delta n)/dT$ and interpreting
the results using MC simulation results, we have been able
to explain why the proportionality between $C_m$ and $d(\Delta n)/dT$
fails at low temperature near the percolation threshold concentration but holds
at higher temperatures.  From the modeling we have done, we have
added to the understanding of the origins of the birefringence in the
experimental system.  We also have shown that the birefringence
technique is well behaved even in a field and, as is well documented\cite{by91},
the critical behavior of the RFIM is faithfully
characterized with the birefringence technique.

This work was funded by Department of Energy Grant No.\ DE-FG03-87ER45324.

\newpage

\begin{figure}[t]
\centerline{\hbox{
\psfig{figure=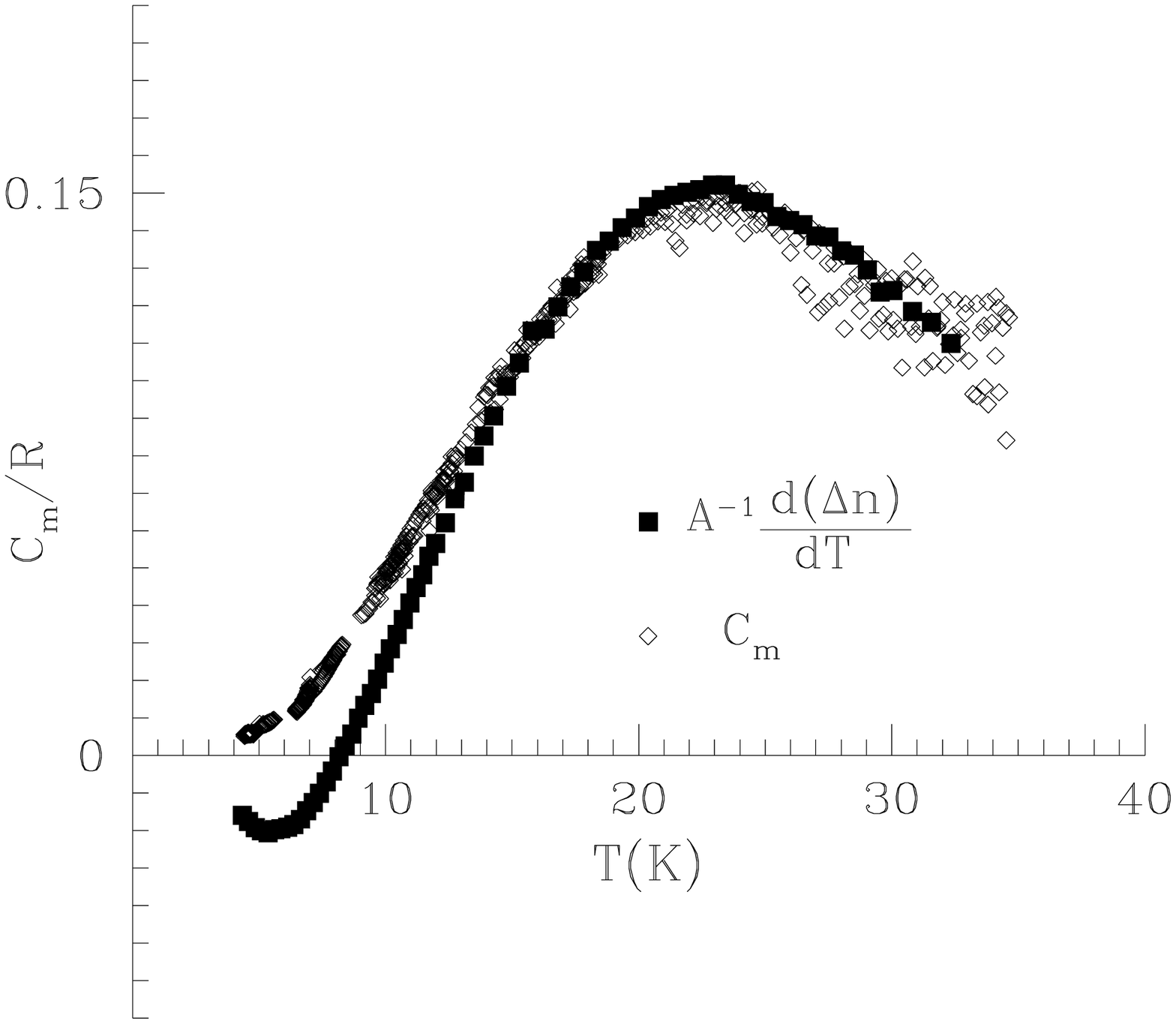,height=5.0in}
}}
\caption
{The magnetic specific heat $C_{m}/R =C_{p}/R - B$, where $B$
is the background, versus $T$ for
$Fe_{0.25}Zn_{0.75}F_2$.  The phonon contribution to the specific heat has
been subtracted as discussed in the text.
Also shown is the temperature derivative of the birefringence $d(\Delta n)/dT$ versus $T$
scaled with the proportionality found at other concentrations.}
\end{figure}

\begin{figure}[t]
\centerline{\hbox{
\psfig{figure=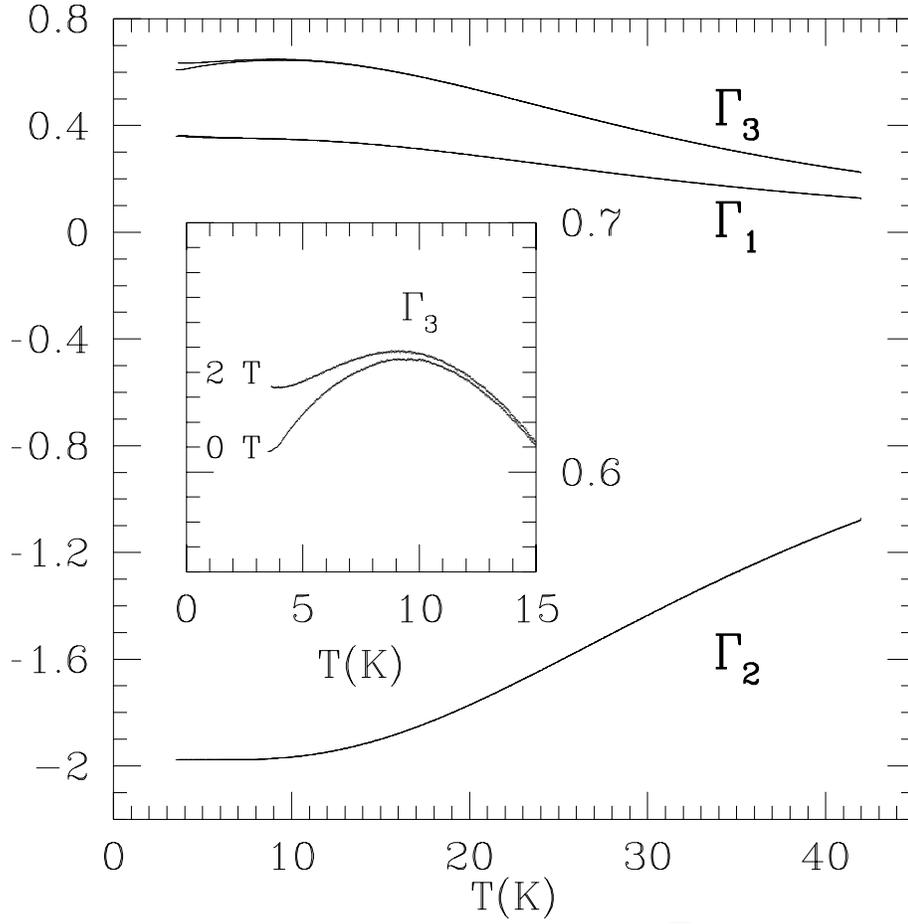,height=5.0in}
}}
\caption
{The first three nearest-neighbor correlation functions
$\Gamma_{l} = \sum_{<i,j>}\epsilon_{i}\epsilon_{j}<S_iS_j>_{l}/N$, where
$N$ is the number of sites, versus $T$
for $Fe_{0.25}Zn_{0.75}F_2$ from MC simulations.
Note the change in sign of the slope in the $H=0$~T curve at $8$~K.}
\end{figure}

\begin{figure}[t]
\centerline{\hbox{
\psfig{figure=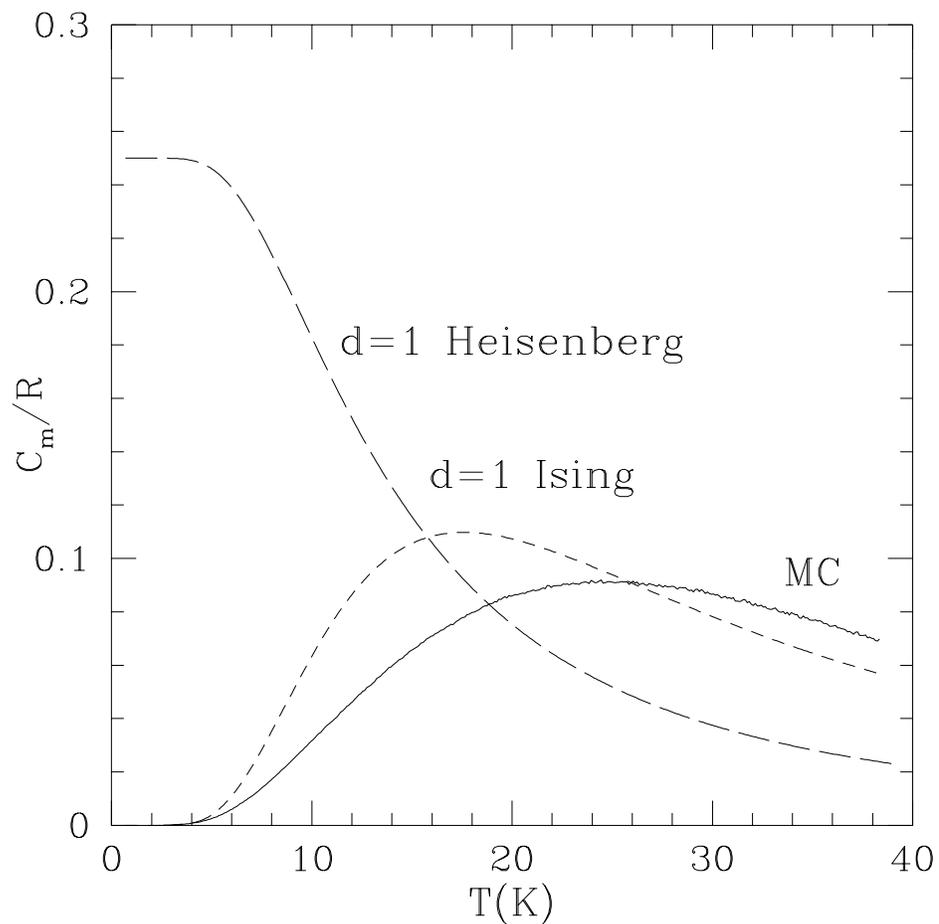,height=5.0in}
}}
\caption
{The magnetic specific heat $C_{m}/R$ versus $T$ for various models. 
The $C_{m}$ per spin predictions have been divided by four in anticipation of a comparison with 
$Fe_{0.25}Zn_{0.75}F_2$, where the molar specific heat is per Avogadro's number 
of molecules, which includes the zinc and is four times larger than the number 
of spins.}
\end{figure}

\begin{figure}[t]
\centerline{\hbox{
\psfig{figure=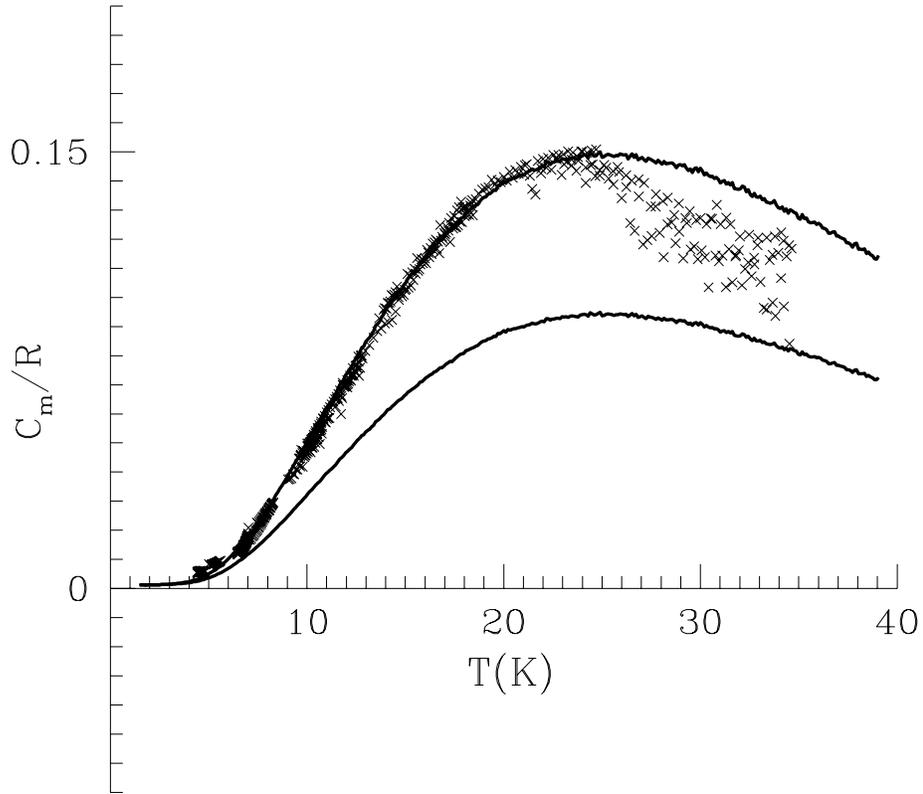,height=5.0in}
}}
\caption
{The magnetic specific heat $C_{m}/R=C_{p}/R - B$,
where $B$ is the background, versus $T$ for
$Fe_{0.25}Zn_{0.75}F_2$.  The phonon contribution to the specific heat has
been subtracted as discussed in the text.  The lower thin line is a fit
using only the second nearest-neighbor correlation function from
MC simulations. The MC results are then scaled by a factor of
1.5 to obtain a good fit to the experimental data at low temperature.}
\end{figure}

\begin{figure}[t]
\centerline{\hbox{
\psfig{figure=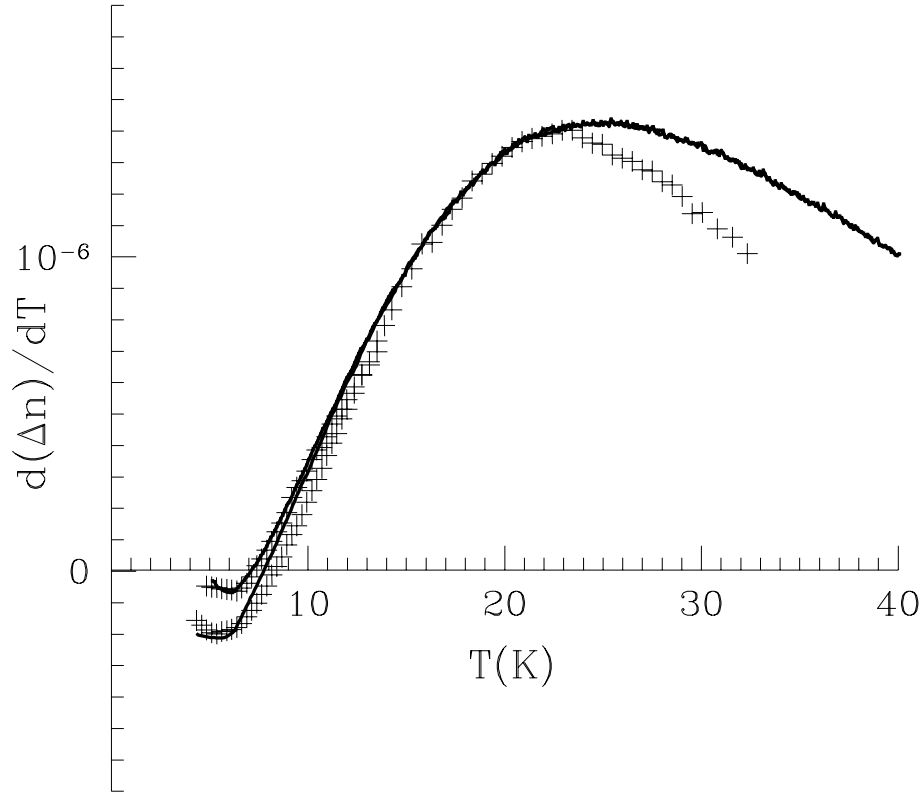,height=5.0in}
}}
\caption
{The temperature derivative of the birefringence $d(\Delta n)/dT$ versus
$T$ for $Fe_{0.25}Zn_{0.75}F_2$.  Note the change in sign in this $H=0$~T
curve at $8$~K.  The curve through the $H=0$ data is a fit using
only the third-nearest-neighbor correlation function from MC
simulations with the coefficient $I_3$ and then scaled by the same factor
of 1.5 used to fit the $C_m$ data in Fig.\ 4.  The curve through the $H=2$~T
data is the corresponding fit using the third-nearest-neighbor
correlation from the $H=2$~T MC simulation.}
\end{figure}

\end{document}